\documentclass[10pt,aps,prl,preprintnumbers,superscriptaddress,showpacs,
twocolumn,floatfix,nofootinbib]{revtex4-1}
\usepackage{graphicx}
\usepackage{amsmath}
\usepackage{slashed}

\newcommand{\bfm}[1]{\mbox{\boldmath$#1$}}

\newcommand{\gsim}{\;\rlap{\lower 3.5 pt \hbox{$\mathchar \sim$}} \raise 1pt
\hbox {$>$}\;}
\newcommand{\lsim}{\;\rlap{\lower 3.5 pt \hbox{$\mathchar \sim$}} \raise 1pt
\hbox {$<$}\;}

\begin{document}

\title{\boldmath
What becomes of vortices when they grow giant
\unboldmath}
\author{Alexander A. Penin}
\email[]{penin@ualberta.ca}
\affiliation{Department of Physics, University of Alberta,
Edmonton, Alberta T6G~2J1, Canada}
\author{Quinten Weller}
\email[]{qweller@ualberta.ca}
\affiliation{Department of Physics, University of Alberta,
Edmonton, Alberta T6G~2J1, Canada}

\begin{abstract}
We discuss vortex solutions of the abelian Higgs model in the limit
of large winding number $n$. We suggest a framework where  a
topological quantum number $n$  is  associated with a ratio of
dynamical scales and a systematic expansion in inverse powers of $n$
is then derived  in the spirit of  effective field theory. The
general asymptotic form of {\it giant} vortices is obtained. For
critical coupling the axially symmetric  vortices become {\it
integrable} in the large-$n$  limit  and we present the corresponding
analytic solution.  The method provides simple asymptotic formulae
for the vortex shape and parameters with accuracy that can be
systematically improved, and can be applied  to topological solitons
of other models. After including the next-to-leading terms the
approximation works remarkably well down to $n=1$.
\end{abstract}
\preprint{ALBERTA-THY-03-20}

\maketitle

Vortices, string-like solutions in   theories with spontaneously
broken gauge symmetry, were originally discovered in the context of
superconductivity \cite{Abrikosov:1956sx} and QCD confinement
\cite{Nielsen:1973cs}. They play a crucial role in many physical
concepts  from cosmic strings \cite{Hindmarsh:1994re} to  mirror
symmetry and dualities of supersymmetric models \cite{Tong:2005un}.
{\it Giant} vortices are observed experimentally in a variety  of
quantum condensed matter systems
\cite{Marston:1977,Engels:2003,Cren:2011}. Corresponding winding
numbers  range  from $n=4$ in  mesoscopic superconductors
\cite{Cren:2011} through $n=60$ in Bose-Einstein condensates of cold
atoms \cite{Engels:2003} and up to  $n=365$ in superfluid $^4$He
\cite{Marston:1977}. Thus, it is quite  appealing  to identify
characteristic  features and universal properties of vortices in the
limit of large $n$, which is a challenging field theory problem.
Though the vortex equations look deceptively simple, their analytic
solution is not available.   Even for critical coupling  when hidden
supersymmetry reduces the order of the equations
\cite{Bogomolny:1975de} and even for the lowest winding number $n=1$
the solution cannot be found in a closed form \cite{deVega:1976xbp}
in contrast, for example, to the apparently more complex case of
magnetic monopoles \cite{Prasad:1975kr}.  Naively one would expect
that finding analytic solutions of higher topological charge should
be  a bigger challenge. In  general, only a few such solutions are
known in gauge models (see {\it e.g.}
\cite{Witten:1976ck,Prasad:1980hg}). However, with increasing winding
number vortices reveal some remarkable properties
\cite{Bolognesi:2005rj,Bolognesi:2005zr}, which  indicate that in the
large-$n$ limit  the solution  may actually become simpler. In this
Letter we suggest a framework that enables a systematic  expansion
in inverse powers of $n$ and find the asymptotic form of the axially
symmetric {\it giant} vortex  solution. Moreover, for  critical
coupling the field equations become integrable and we present the
corresponding analytic result.

Since an expansion in inverse powers of  a topological charge  may
not be overly intuitive let us first outline its main idea. When the
winding number $n$ grows, the characteristic size of the vortex has
to grow as well  to accommodate the  increasing  magnetic flux. Assuming
a roughly uniform average distribution of the flux inside the vortex
we get an estimate of its radius $\sqrt{n}/e$, where  $e$ is the
gauge charge of the  scalar field.  At the same time a characteristic
distance of the nonlinear interaction is $1/e$. Thus for large $n$
we get a scale hierarchy and the expansion in the corresponding scale
ratio is a standard tool  of the effective field theory approach.
Since we deal with the spatially extended classical solutions it is
more convenient to perform  this expansion in coordinate space at the
level of the equations of motion.

We consider the standard Lagrangian for the abelian Higgs
(Ginzburg-Landau) model of a scalar field with  abelian charge $e$,
quartic  self-coupling $\lambda$, and vacuum expectation value $\eta$
in two dimensions
\begin{equation}
{L}=-{1\over 4} F^{\mu\nu}F_{\mu\nu}
+\left({D^\mu \phi}\right)^\dagger D_\mu \phi
-{\lambda\over 2}\left(\left|\phi\right|^2-\eta^2\right)^2\,,
\label{eq::Lagrange}
\end{equation}
where $D_\mu=\partial_\mu+ieA_\mu$. Vortices are topologically
nontrivial solutions of the Euclidean equations of motion. For
critical coupling $\lambda=e^2$ these reduce to the  first-order
Bogomolny equations \cite{Bogomolny:1975de}
\begin{equation}
\begin{split}
& \left(D_1+iD_2\right)\phi =0\,, \\
& -F_{12}+e\left(\left|\phi\right|^2-1\right)=0\,.
\label{eq::BogomolnyphiF}
\end{split}
\end{equation}
We then study the  axially symmetric solutions  of winding number $n$,
which in polar coordinates can be written as follows
$\phi(r,\theta)=f(r)e^{in\theta}$, $A_\theta=-n a(r)/e$, $A_r=0$. It
is convenient to work with the rescaled dimensionless quantities
$e\eta r\to r$, $f/\eta\to f$, $\lambda/e^2\to\lambda$ so that in
the new variables $e=\eta=1$ and critical coupling corresponds to
$\lambda=1$. Then the Bogomolny equations in terms of the functions
$a(r)$ and $f(r)$ take the following form
\begin{equation}
\begin{split}
& {df\over dr}-{n\over r}(1-a)f  = 0 \,, \\
& {da\over dr}+{r\over n}(f^2-1) = 0\,,
\label{eq::Bogomolnyfa}
\end{split}
\end{equation}
with the boundary conditions $f(0)=a(0)=0$ and $f(\infty)=a(\infty)=1$.
For a given winding number the solution  carries $n$ quanta of
magnetic flux $\Phi= -\int F_{12}{\rm d}^2{\bfm r}=2\pi n$ and the
energy or string tension $T= -\int L{\rm d}^2{\bfm r}=2\pi n\eta^2$.

For large $n$ the field dynamics is essentially different in three
regions: the core,   the boundary layer, and the tail of the vortex.
Below we discuss the specifics of the dynamics and its description
in each region.

\noindent
{\em The vortex core.} For small $r$ the solution of the field
equations gives $f(r)\propto r^n$. This function is exponentially
suppressed at large $n$ for all $r$ smaller than a critical value
which  can be associated with the core boundary. For such $r$ the
contribution of $f$ can be neglected in the equation for $a$ and we
get $a(r)\approx r^2/r_n^2$ with $r_n=\sqrt{2n}$, which in turn can
be used in the equation for $f$. Thus in the core the dynamics is
described by linearized equations in the background field
\begin{equation}
\begin{split}
& {df\over dr}-{n\over r}\left(1-{r^2\over r_n^2}\right)f  = 0 \,, \\
& {da\over dr}-{r\over n} = 0\,.
\label{eq::coreeq}
\end{split}
\end{equation}
Their  solutions read
\begin{equation}
\begin{split}
&f(r)=\exp\left[{n\over 2}\left(\ln\left({r^2\over r^2_n}\right)
- {r^2\over r_n^2}+1-{1\over n}\right)\right],\\
&a(r)={r^2\over r_n^2}\,,
\label{eq::coresol}
\end{split}
\end{equation}
where the form of the integration constant in the first line is
determined by matching conditions explained below. For
$r_n-r={\cal O}(1)$ we have  $n(1-a)/r={\cal O}(1)$ and the equation
for $f$ becomes independent of $n$. Hence the approximation
Eq.~(\ref{eq::coreeq}) is not applicable anymore, the nonlinear
effects become crucial, and  we enter the boundary layer.  Note that
the magnetic flux and energy density for Eq.~(\ref{eq::coresol}) are
approximately $1$ and $\eta^2$, respectively, so that  the core
accommodates essentially all the vortex flux and energy and  we can
identify $r_n$ with the vortex radius.

\begin{figure}[t]
\begin{center}
\includegraphics[width=8cm]{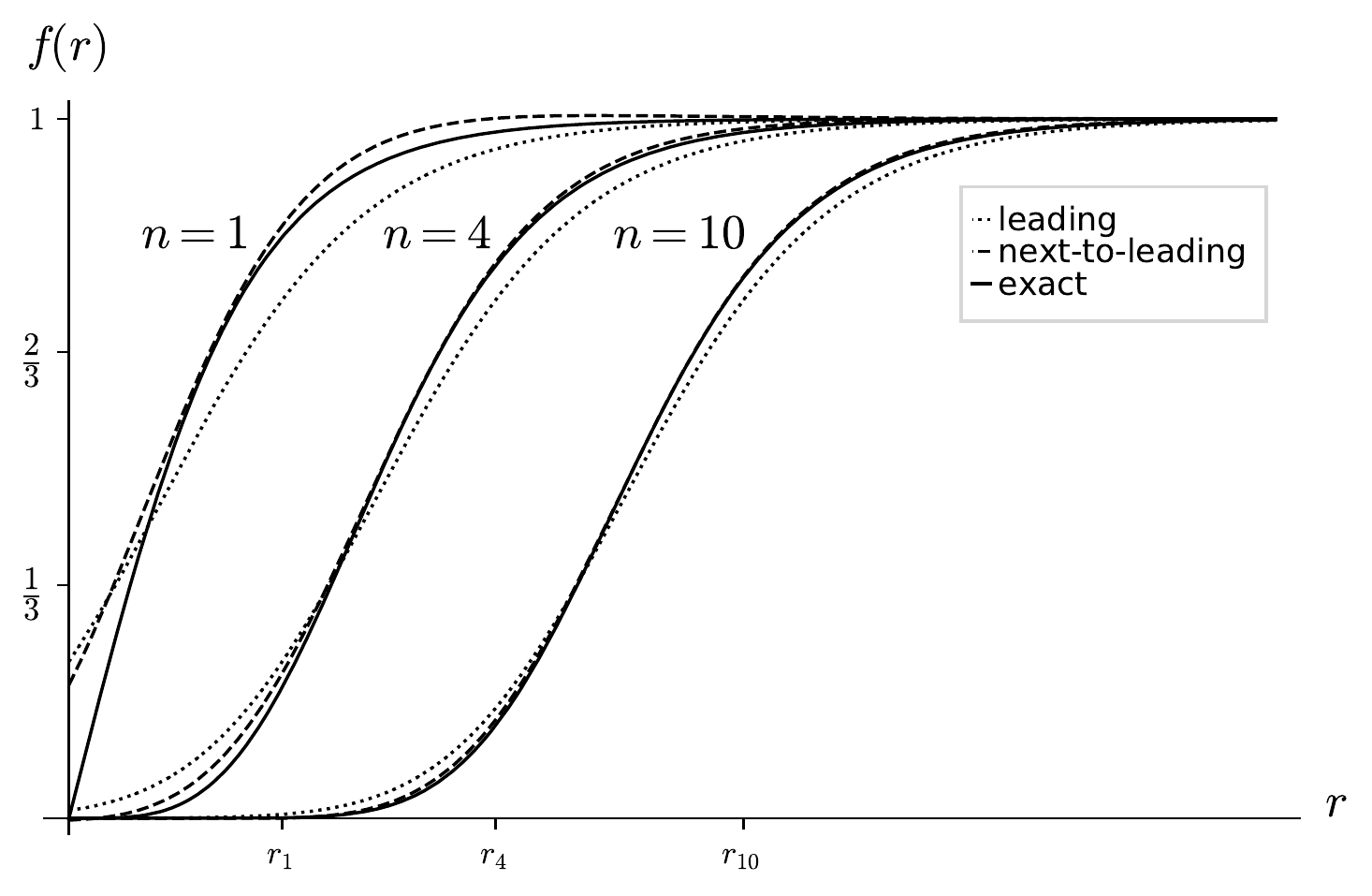}
\end{center}
\caption{\label{fig::1}   The  numerical  solution of the exact
critical vortex equations for the scalar field $f(r)$ (solid lines),
the leading asymptotic solution $e^{w(r-r_n)}$ (dotted lines), and
the next-to-leading approximation  (dashed lines) for different
winding numbers $n$.}
\end{figure}

\noindent
{\em The  boundary layer.} In this region the field dynamics is
ultimately nonlinear. However, it crucially simplifies  for large $n$.
To see this we introduce a new radial coordinate $x=r-r_n$ so that in
the boundary layer $x = {\cal O}(1)$ and the expansion in $x/r_n$
converts into an expansion in $1/\sqrt{n}$. In the leading order in
$x$ Eq.~(\ref{eq::Bogomolnyfa}) reduces to  a system of
$n$-independent field equations with constant coefficients
\begin{equation}
\begin{split}
& w'+\gamma= 0 \,, \\
& \gamma' -1+e^{2w}=0\,,
\label{eq::boundeq}
\end{split}
\end{equation}
where  $w(x)=\ln f(r_n+x)$,
$\gamma(x)={n}\left(a(r_n+x)-1\right)/r_n$, and prime stands for a
derivative in $x$.  The system can be resolved for $w$ which results
in  a second-order equation
\begin{equation}
\begin{split}
& w''+1-e^{2w}= 0\,.
\label{eq::weq}
\end{split}
\end{equation}

\begin{figure}[t]
\begin{center}
\includegraphics[width=8cm]{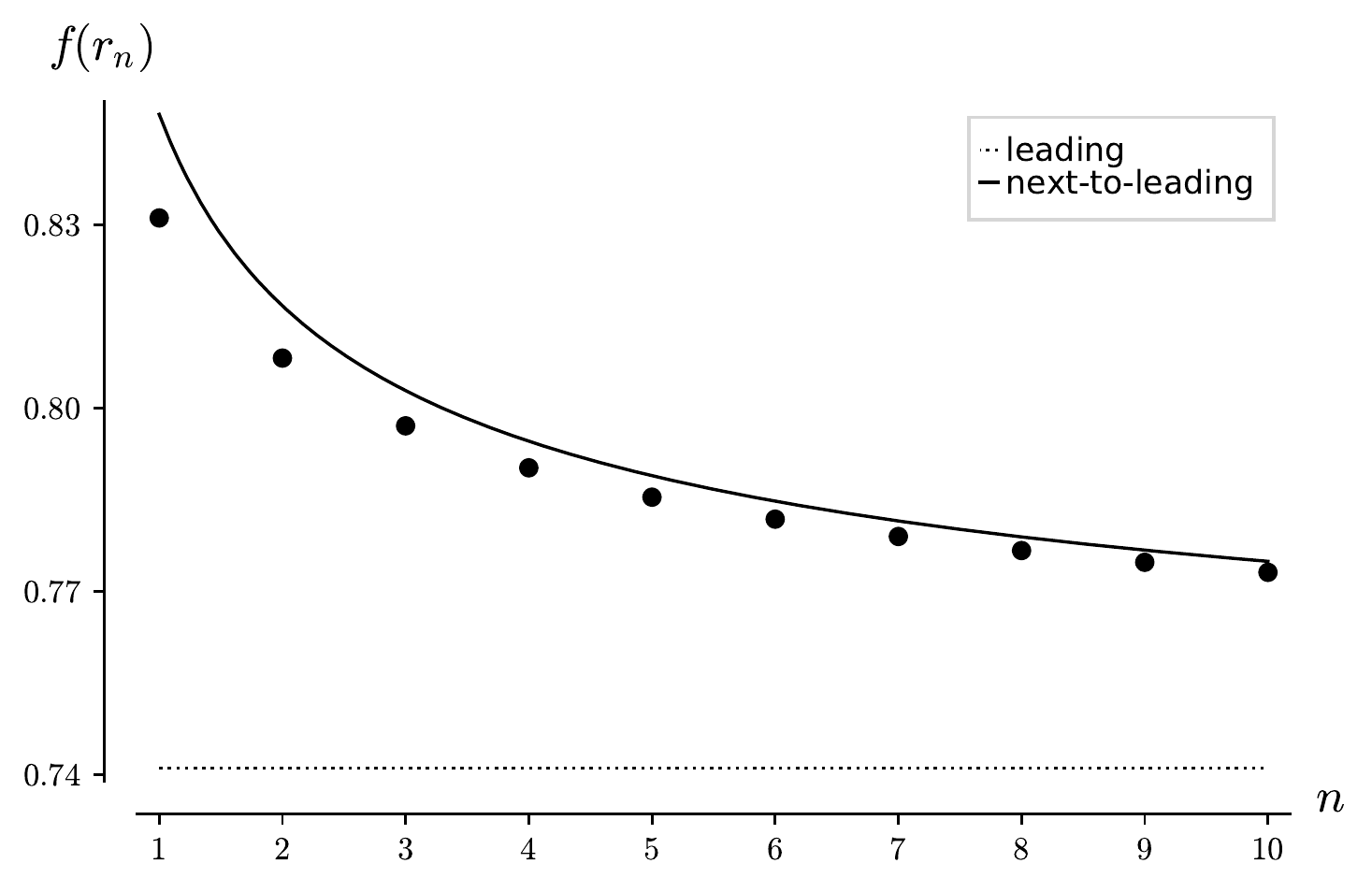}
\end{center}
\caption{\label{fig::2}   The  values $f(r_n)$ obtained from the
numerical solution of the exact critical vortex equations (black
dots), the asymptotic value $f(r_n)=e^{w_0}$ (dotted line), and the
next-to-leading approximation (solid line) as functions of the
winding number $n$.}
\end{figure}

\noindent
This equation  has a first integral $I=w'^2-e^{2 w}+2w$ with $I=-1$
corresponding to the boundary condition $w(\infty)=0$. Thus
Eq.~(\ref{eq::boundeq}) can be solved in quadratures with the result
\begin{equation}
\begin{split}
& \int^{w(x)}_{w_0} {{\rm d}w \over (e^{2 w}-2w-1)^{1/2}}=x  \,,\\
& \gamma(x) =  -(e^{2 w(x)}-2w(x)-1)^{1/2}\,,
\label{eq::boundsol}
\end{split}
\end{equation}
where $w_0=w(0)$  is the second  integration constant.  It is
determined by the boundary condition $w'(x)\sim -x$ at $x\to-\infty$,
which ensures that Eq.~(\ref{eq::boundsol}) can be matched to the
core solution. This gives a new transcendental constant
\begin{equation}
w_0=-0.2997174398\ldots\,,
\label{eq::wellerconst}
\end{equation}
which determines a unique asymptotic solution in the boundary layer.
It has the Taylor expansion  $w(x)=\sum_{m=0}^\infty w_mx^m$ where
$w_1=(e^{2w_0}-2w_0-1)^{1/2}$  and the higher order coefficients can
be obtained recursively.  The asymptotic behavior of the function at
$x\to\infty$ reads
\begin{equation}
\begin{split}
&w(x) \sim w_\infty e^{-\sqrt{2}x}  \,,\\
& w(-x) \sim -{x^2\over 2}-{1\over 2}+\ldots\,,
\label{eq::wasym}
\end{split}
\end{equation}
\noindent
where $w_\infty=w_0\exp[\int_{w_0}^0(\sqrt{2}/(e^{2w} -2w-1)^{1/2}+{1/
w}){\rm d}w]$. By using Eq.~(\ref{eq::wasym}) it is straightforward
to verify that up to corrections suppressed at large $n$ the boundary
layer solution Eq.~(\ref{eq::boundsol}) coincides with the core
solution Eq.~(\ref{eq::coresol}) in the matching region $1\ll r_n-r
\ll r_n$, where both approximations are valid. This is a rather
nontrivial result since Eq.~(\ref{eq::coresol}) does depend on $n$.

\begin{figure}[t]
\begin{center}
\includegraphics[width=8cm]{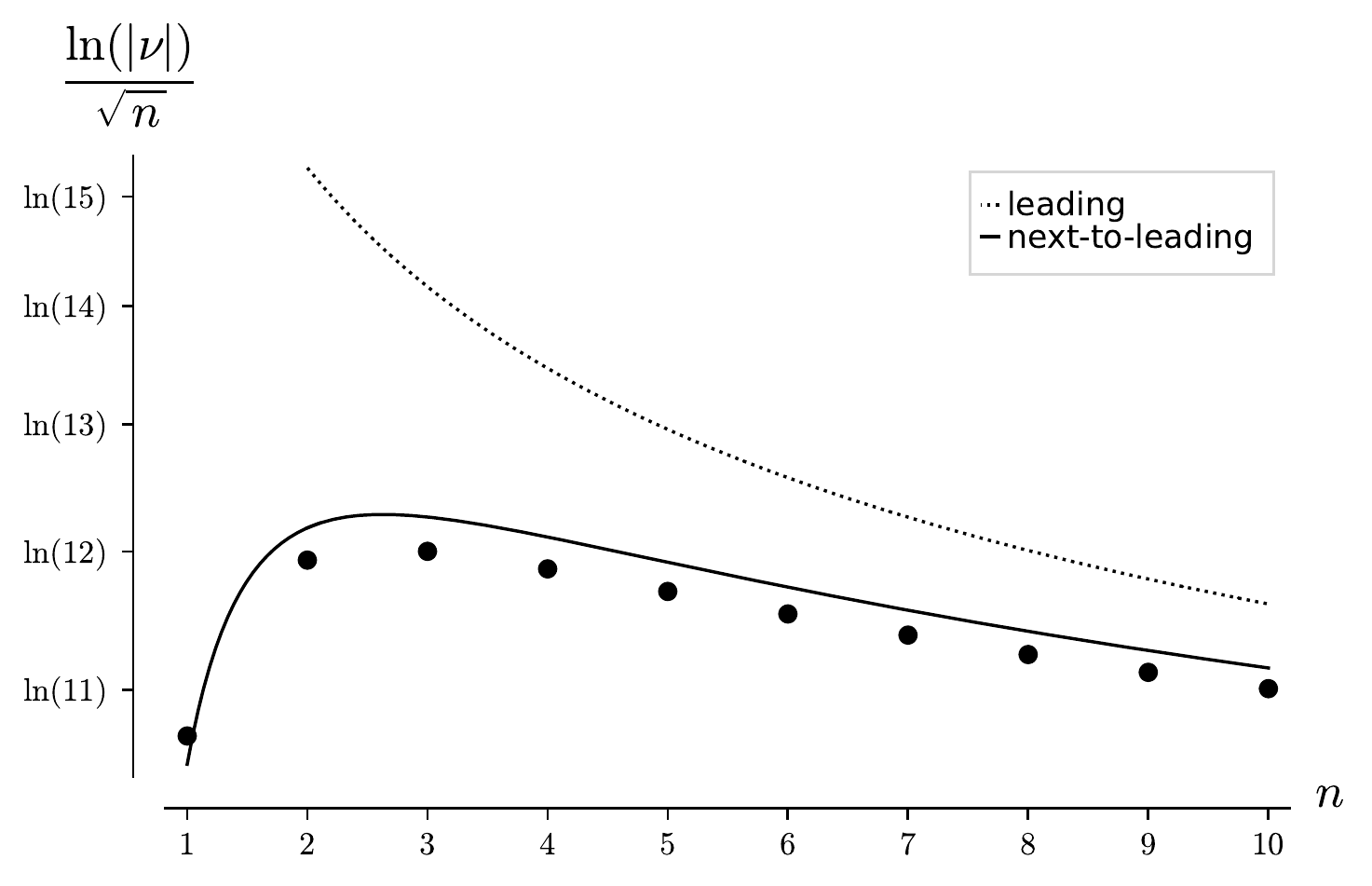}
\end{center}
\caption{\label{fig::3} Scalar charge/magnetic dipole moment  $\nu$
of a critical vortex obtained from the numerical  solution of the
exact vortex equations (black dots), the asymptotic result
Eq.~(\ref{eq::charge})  (dotted line), and the next-to-leading
approximation (solid line) as functions of the winding number $n$.}
\end{figure}

\noindent
{\em The vortex tail.} For $(r-r_n)/r_n={\cal O}(1)$  the boundary
layer approximation breaks down and the coordinate dependence of the
field equation coefficients should be restored. However, the deviation
of the fields from the vacuum configuration is now exponentially
small so the field  equations linearize. The solution of the
linearized theory is well known and reads
\begin{equation}
\begin{split}
& f(r) \sim 1+ {\nu\over 2\pi}K_0(\sqrt{2}r)+\ldots\,, \\
&a(r) \sim 1+ {\mu\over 2\pi}{\sqrt{2}r}K_1(\sqrt{2}r)+\ldots\,,
\label{eq::tailsol}
\end{split}
\end{equation}
where $K_m(z)$ is the $m$th modified Bessel function. It describes
the field of a point-like source of scalar charge $\nu$ and magnetic
dipole moment $\mu$ with $\nu=\mu$ for critical coupling.
Eqs.~(\ref{eq::boundsol}) and (\ref{eq::tailsol}) should coincide in
the second matching region $1\ll r-r_n \ll r_n$, which yields
\begin{equation}
\nu=4w_\infty \sqrt{\pi}e^{2\sqrt{n}+\ln(n)/4}\,.
\label{eq::charge}
\end{equation}
Thus vortex scalar charge and  magnetic dipole moment  grow
exponentially with the winding number.

Calculation of the higher order terms  of the expansion in
$1/\sqrt{n}$ is rather straightforward. Writing down the leading
corrections to the asymptotic solutions  $w$ and $\gamma$ as  $\delta
w/\sqrt{2n}$ and  $\delta \gamma/\sqrt{2n}$, respectively, we get
\begin{eqnarray}
&&\delta w(x)= C w'(x) +\int_{0}^{x}{w'(x)\over w'^2(z)} \int^{\infty}_{z}w'^2(y)
{\rm d}y {\rm d}z\,, \nonumber\\
&&  \delta \gamma(x) = -{x}w'(x)-\delta w'(x)\,,
\label{eq::nlo}
\end{eqnarray}
where $C=\int_{-\infty}^{0} [ z/3 + \int^{\infty}_{z}w'^2(y)/w'^2(z) dy ] dz$.

\begin{figure}[t]
\begin{center}
\includegraphics[width=8cm]{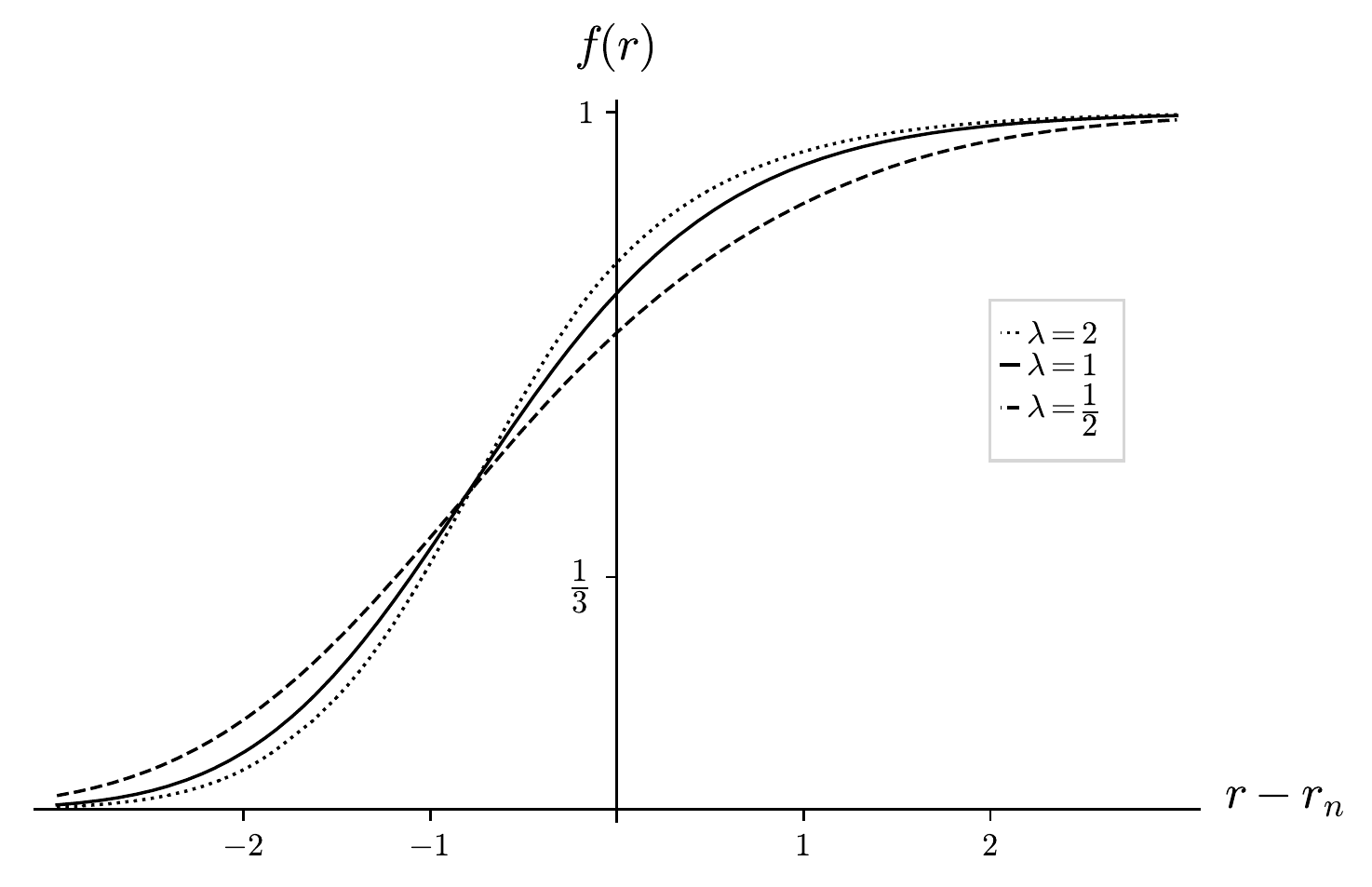}
\end{center}
\caption{\label{fig::4}   The  asymptotic profiles of the scalar
field $f(r)$ obtained by  numerical  solution of the effective vortex
equations Eq.~(\ref{eq::efffieldeq}) for different values of scalar
self-coupling as functions of $r-r_n$.}
\end{figure}

\noindent

Let us now consider noncritical coupling $\lambda\ne 1$. In this
case the order of the field equations  cannot be reduced and
they read
\begin{equation}
\begin{split}
&{1\over r}{d\over dr}\left(r{df\over dr}\right)-
 \left[\lambda(f^2-1)+{n^2\over r^2}(1-a)^2\right]f  = 0 \,,\\
& r{d\over dr}\left({1\over r}{da\over dr}\right)+{2}(1-a)f^2 = 0\,.
\label{eq::fieldeqfa}
\end{split}
\end{equation}
Nevertheless, the general structure of the solution  is quite similar
to the critical case. Inside the core the contribution of the scalar
potential to Eq.~(\ref{eq::fieldeqfa}) is suppressed by $r^2/n^2$.
Hence the core dynamics is not sensitive to  $\lambda$ and the core
solution is given by  Eq.~(\ref{eq::coresol}) up to the value of the
integration constants which do depend on $\lambda$ through the
matching to the nonlinear boundary layer solution. In particular the
vortex size $r_n$ is  determined by the region where the two terms in
the square brackets of Eq.~(\ref{eq::fieldeqfa}) become  comparable
and the core approximation breaks down,  which gives the leading
order result $r_n=\sqrt{2n}/\lambda^{1/4}$. Note that the
approximately constant energy density in the core is now
$\lambda\eta^2$ so that the  total vortex energy in the large-$n$
limit  is $T=2\pi\sqrt{\lambda}n\eta^2$. This agrees with the
``wall-vortex'' conjecture and numerical results for very large $n$
of Refs.~\cite{Bolognesi:2005rj,Bolognesi:2005zr}. In the tail
solution, Eq.~(\ref{eq::tailsol}), the argument of $K_0$ gets an
additional factor of $\sqrt{\lambda}$ to account for the variation of
the scalar field mass, while the scalar charge and the magnetic
dipole moment are not equal anymore and have different leading
behavior  at $n\to \infty$
\begin{equation}
\begin{split}
&|\nu|\sim e^{2\sqrt{n}\,\lambda^{1/4}+\ldots} \,,\qquad
|\mu|\sim e^{2\sqrt{n}/\lambda^{1/4}+\ldots}\,.
\label{eq::chargelam}
\end{split}
\end{equation}
More accurately these parameters as well as the normalization of the
scalar field in the core solution are determined by matching to the
boundary layer solution. In the boundary layer by expanding in
$x/r_n$ we get a system of $n$-independent equations with constant
coefficients
\begin{equation}
\begin{split}
&{f''}-
 \left[{\lambda}\left({f^2}-1\right)+{\gamma^2}\right]f  = 0 \,,\\
& \gamma''-{2}\gamma{f^2} = 0\,,
\label{eq::efffieldeq}
\end{split}
\end{equation}
with the boundary condition $\gamma(x)\sim \sqrt{\lambda}x$ at
$x\to-\infty$. For $\lambda=1$ the proper  solution is given by
Eq.~(\ref{eq::boundsol}) and for any given $\lambda\ne 1$ it can be
found numerically.

Finally we briefly discuss a simpler but quite interesting case of
the  vortices in Bose-Einstein condensate of a neutral  scalar field.
The corresponding vortex equation is obtained from the first line of
Eq.~(\ref{eq::fieldeqfa}) by setting $a=0$ and $\lambda =1$ (see,
{\it e.g.} \cite{Landau}). Now the dynamics in the core of  radius
$r_n=n$ is described by a linear differential equation, while in the
tail region $ r_n+\delta\lsim r$ with $\delta\propto n^{1/3}$  the
derivative term  is suppressed and  the field equation becomes
{\it algebraic} at $n\to\infty$. In contrast to the charged case the
boundary layer does not form and the core and tail solutions can be
polynomially  matched over the interval $r_n\lsim r\lsim  r_n+\delta$.
This yields the asymptotic solution
\begin{equation}
f(r)=\left\{
\begin{array}{l}
 C'J_n(r)\,,  \quad r\le r_n\,,  \\[3pt]
 \sqrt{\delta/ 2n}
 \left(1+{(r-r_n)/\delta}\right),\,
 r_n< r< r_n+\delta\,,  \\[3pt]
 \sqrt{1-{n^2/  r^2}}\,,
 \quad r_n+\delta\le r\,,
\end{array}
\right.
\label{eq::BECsol}
\end{equation}
where $J_n(r)$ is the $n$th Bessel function,
$C'={3^{1/4}\pi^{1/2}\over 2^{1/2}}$, $\delta={2^{2/3}\pi n^{1/3}\over
3^{5/6}\Gamma^2(2/3)}$, and $\Gamma(z)$ is the Euler gamma-function.
The vortex energy now is $T=\pi \eta^2 n^2$, where a half of the
contribution comes from the vortex tail. Curiously, for $\eta=1$
it is given by  the area of a  circle of radius $n$ while the critical
vortex  energy is equal to the corresponding circumference.  The
corrections to Eq.~(\ref{eq::BECsol}) are given by a series in
$1/n^{1/3}$ and will be published  elsewhere.

\begin{figure}[t]
\begin{center}
\includegraphics[width=8cm]{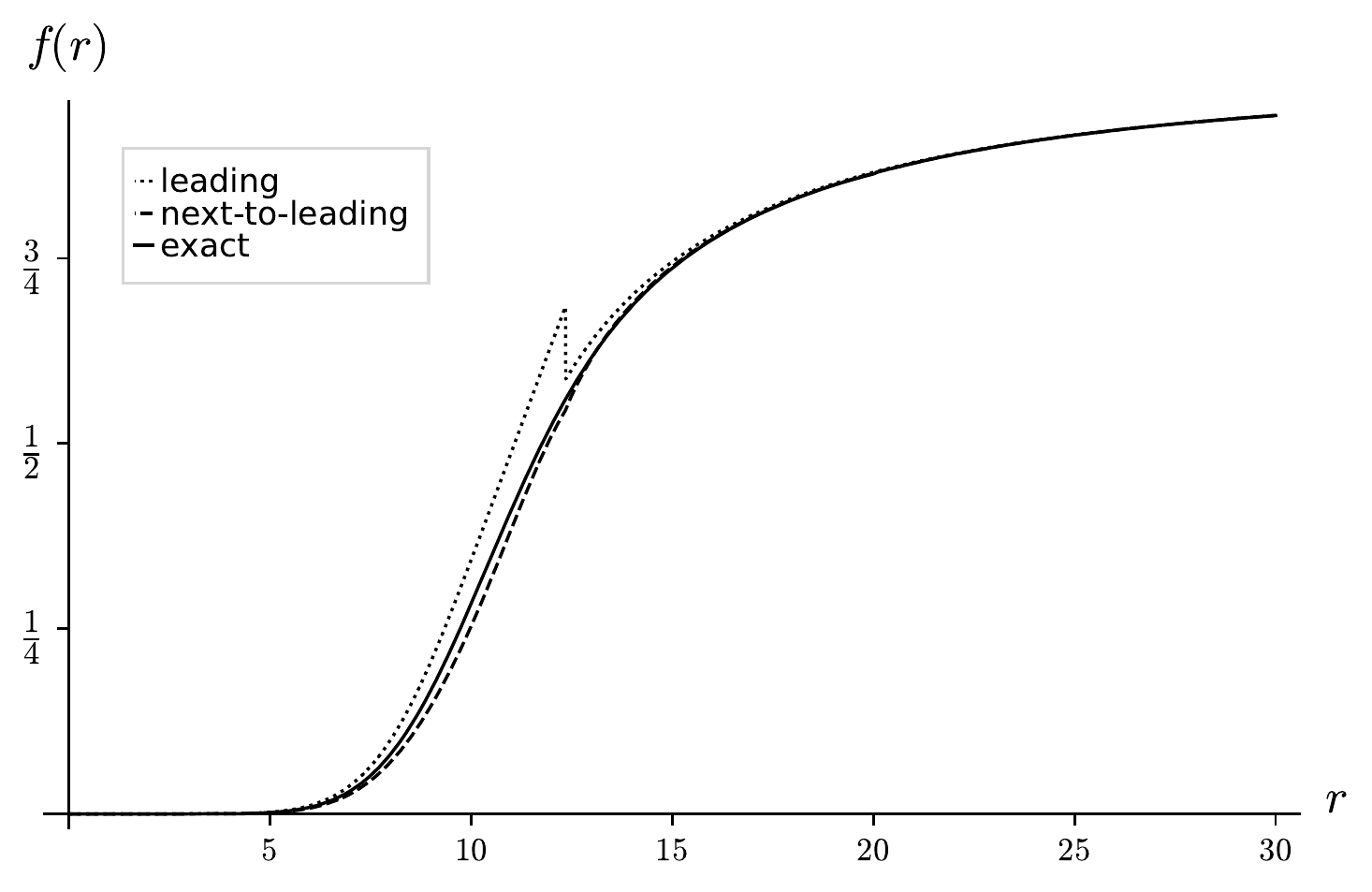}
\end{center}
\caption{\label{fig::5}   The  numerical  solution of the exact
vortex equations for a neutral scalar field $f(r)$ (solid line),
the leading asymptotic solution Eq.~(\ref{eq::BECsol})  (dotted
line), and the next-to-leading approximation  (dashed line) for
$n=10$.}
\end{figure}

The results of numerical analysis of the large-$n$ approximation are
presented in Figs.~\ref{fig::1}-\ref{fig::5}. In Fig.~\ref{fig::1}
the leading asymptotic result $e^{w(r-r_n)}$ and the next-to-leading
approximation which incorporates the ${\cal O}(1/\sqrt{n})$ terms are
plotted against numerical solutions of the exact field equations for
$f(r)$ with $\lambda=1$, $n=1,\,4,\,10$. In Figs.~\ref{fig::2} and
\ref{fig::3} the exact numerical values of $f(r_n)$ and $\nu$, the
natural characteristics of the vortex solution, are plotted against
the asymptotic leading and next-to-leading results for $\lambda=1$,
$n\le 10$. The expansion reveals an impressive convergence and the
next-to-leading  approximation works reasonably well even for $n=1$.
For completeness we present the result for the asymptotic profile of
the boundary layer solution for the scalar field with $\lambda=1/2,\,
1,\, 2$ in Fig.~\ref{fig::4}. The numerical results for a neutral
scalar field vortex with $n=10$ are given in Fig.~\ref{fig::5}

To summarize, we have elaborated a method of  expansion in  inverse
powers of a topological quantum number.  The method is quite general
and can be applied to the study of topological solitons in a theory
where the corresponding  quantum number can be associated with a
ratio of dynamical scales, {\it e.g.} to the multi-monopole solutions
in Yang-Mills Higgs  model, where only the case of vanishing scalar
potential has been solved so far. When applied to axially symmetric
vortices with large winding number $n$ the expansion is in powers of
$1/n^{1/\alpha}$ with $\alpha=2$ for the charged and  $\alpha=3$ for
the neutral scalar field.  In the large-$n$ limit the complex
nonlinear vortex dynamics unravels. In particular,  the field
equations become integrable for critical coupling and reduce to
an algebraic one for a neutral Bose-Einstein condensate. This yields
simple asymptotic formulae for the shape and parameters capturing the
main features of the {\it giant} vortices. The accuracy of the
asymptotic result can be systematically improved and already after
including the leading corrections the approximation works remarkably
well all the way down to very low $n$.

\acknowledgments
A.P.  is grateful to Joseph Maciejko  for useful communications. The
work of  A.P. was supported in part by NSERC and the Perimeter
Institute for Theoretical Physics. The work of  Q.W. was supported
through the NSERC USRA program.


\end{document}